\documentclass[11pt, a4paper]{amsart}
\usepackage{amsmath}
\usepackage{amsxtra}
\usepackage{amscd}
\usepackage{amsthm}
\usepackage{amsfonts}
\usepackage{amssymb}
\usepackage {pstricks}
\usepackage{pstricks,pst-node}

\theoremstyle{definition}

\theoremstyle{remark}

\numberwithin{equation}{section}

\newcommand{\R}{{\mathbb R}}

\newcommand{\barh}{{\bar h}}
\newcommand{\cA}{{\mathcal A}}

\newcommand{\g}{{{\bf g}}}

\begin{document}


\title[Quantum deformations of spacetime]
{Quantum deformations of Schwarzschild and Schwarzschild-de Sitter
spacetimes}

\author{Ding Wang}
\address{Institute of Mathematics,
Academy of Mathematics and Systems Science, Chinese Academy of
Sciences, Beijing 100190, China}
\email{wangding@amss.ac.cn}

\author{R. B. Zhang}
\address{School of Mathematics
and Statistics, University of Sydney, Sydney, NSW 2006, Australia}
\email{rzhang@maths.usyd.edu.au}

\author{Xiao  Zhang}
\address{Institute of Mathematics,
Academy of Mathematics and Systems Science, Chinese Academy of
Sciences, Beijing 100190, China}
\email{xzhang@amss.ac.cn}

\begin{abstract}
A quantum Schwarzschild spacetime and a quantum Schwarzschild-de
Sitter spacetime with cosmological constant $\Lambda$ are
constructed within the framework of a noncommutative Riemannian
geometry developed in an earlier publication. The metrics and
curvatures of the quantum Schwarzschild spacetime and the quantum
Schwarzschild-de Sitter spacetime are computed. It is shown that up
to the second order in the deformation parameter, the quantum
spacetimes are solutions of a noncommutative Einstein equation.
\end{abstract}
\maketitle

%
%
\section{Introduction}\label{introd}

There have been intensive activities studying noncommutative
analogues of general relativity. The studies are largely motivated
by the widely held belief that the usual notion of spacetime as a
pseudo-Riemannian manifold needs to be modified at the Planck scale.
Indeed it was demonstrated in \cite{DFR} that at the Planck scale,
the coordinates describing spacetime points satisfied certain
nontrivial commutation relations analogous to those appearing in
quantum mechanics. This suggests the relevance of noncommutative
geometry to Planck scale physics.

A large volume of work has been done on noncommutative analogues of
black holes \cite{R1, R2, R3, R4} by using a variety of physically
motivated methods and incorporating different physical intuitions.
The papers in \cite{R2} assumed a Gaussian distribution for matter
in Einstein gravity and analyzed the integrated effect of
noncommutativity as corrections to the classical Schwarzschild black
hole. Analysis in a similar vein was carried out in \cite{R3}. In
\cite{R4},  Chaichian and collaborators investigated corrections to
physical quantities of black holes arising from the noncommutativity
of spacetime itself by using a gauge theoretical formulation.
General relativity on a noncommutative spacetime is regarded in
\cite{R4} as a nonconmmutative gauge theory of a deformed Lorentzian
algebra analogous to the classical picture of Utiyama and Kibble. A
methodology akin to this is also adopted in \cite{R6, AMV, R7}.
Particularly noteworthy is the study in \cite{AMV}, which showed
that a deformation of general relativity arising from \cite{R7}
differed qualitatively from the low energy limit of string theory.
Very recently, Buric and Madore in \cite{BM} (see also \cite{R10})
explored a possible moving frame formalism for a noncommutative
geometry on the Moyal space as the first step toward setting up a
framework for studying Schwarzschild or Schwarzschild Sitter black
holes.

In \cite{CTZZ} a theory of noncommutative Riemannian geometry over
the Moyal algebra was developed by studying noncommutative surfaces
embedded in higher dimensions. This theory makes use of Nash's
isometric embedding theorem \cite{N} and its generalisations
\cite{C}. It retains key notions of usual Riemannian geometry, such
as the metric and curvature, which are essential for describing
gravity. Other features of the theory of \cite{CTZZ} are its
simplicity and transparent consistency, which render the theory
particularly amenable to explicit computations. As we shall see in
this paper, computations within the theory are no more difficult
than that in usual Riemannian geometry. Furthermore, it is possible
to apply the formalism of \cite{CTZZ} to obtain a theory of
$\kappa$-deformed \cite{LNR} noncommutative geometry. The
$\kappa$-deformed spacetimes (see \cite{L} for references)  are
another class of noncommutative spacetimes much studied in the
literature.

The present paper applies the theory of noncommutative Riemannian
geometry developed in \cite{CTZZ} to investigate quantum aspects of
gravity from a mathematical point of view. Specifically, we
construct quantum deformations of the Schwarzschild spacetime and
the Schwarzschild-de Sitter spacetime in the framework of
\cite{CTZZ} and investigate the physical properties of such
noncommutative spacetimes. The key results on the quantum
Schwarzschild spacetime are equations \eqref{deformed-Sch} and
\eqref{RTheta}, which respectively give the metric and the Ricci
curvature.  The metric and the Ricci curvature of the quantum
Schwarzschild-de Sitter spacetime are respectively given by
equations \eqref{deformed-Sch-dS} and \eqref{RTheta-dS}. The Hawking
temperature and entropy of the quantum Schwarzschild black hole are
investigated, and a quantum correction to the entropy-area law is
observed (see \eqref{entropy}).

The noncommutative analogue of the Einstein equation in the vacuum
proposed in \cite{CTZZ} is generalised to include matter (equation
\eqref{Einstein}). We show that the quantum Schwarzschild spacetime
and quantum Schwarzschild-de Sitter spacetime are solutions of
\eqref{Einstein} in the vacuum exact to the first order in the
deformation parameter. However, higher order terms appear to require
matters sources. It will be interesting to investigate the physical
origin and implications of the source terms.

The organisation of the paper is as follows. In Section
\ref{spacetimes}, we briefly review the noncommutative Riemannian
geometry developed in Ref. \cite{CTZZ} in the light of Nash's
isometric embedding theorem \cite{N} and its generalisation to
pesudo-Riemannian manifolds \cite{C}. In Section
\ref{schwarzschild}, we present two constructions of a quantum
Schwarzschild spacetime and study its noncommutative geometry.  The
Hawking temperature and entropy of the quantum Schwarzschild black
hole are also analyzed. Section \ref{schwarzschild-dS} constructs
the quantum Schwarzschild-de Sitter spacetime and studies its
noncommutative geometry. Section \ref{conclusion} concludes the
paper with some brief comments on the results obtained.

\section{Local quantum deformation of spacetimes}\label{spacetimes}

Let $(N^{1,n-1}, g)$ be an $n$-dimensional Lorentzian manifold whose
metric $g$ has signature $(-1,1,\cdots,1)$. By results of
\cite{C},  which extends Nash's isometric embedding to
pesudo-Riemannian manifolds, there exist positive integers $p$, $q$
and a set of smooth function $X^1, \cdots, X^p, X^{p+1},\cdots,
X^{p+q}$ on $N^{1,n-1}$ such that
 \begin{eqnarray}
g=-\big(dX^1\big)^2-\cdots
-\big(dX^p\big)^2+\big(dX^{p+1}\big)^2+\cdots+\big(dX^{p+q}\big)^2.
 \end{eqnarray}

Let $U$ be a coordinate chart of $N^{1,n-1}$ with natural
coordinates $\{x^0, x^1, \cdots, x^n\}$. Following the mathematical
convention, we allow the possibility that some of the coordinates
are dimensionless instead of having the dimension of length or time.
Let $\barh$ be a real indeterminate, and denote by $\R[[\barh]]$ the
ring of formal power series in $\barh$. Here $\barh$ is taken to be
dimensionless. Physically one may regard $\barh$ as the ratio of the
standard model mass scale and the Planck mass. Let $\cA$ be the set
of formal power series in $\barh$ with coefficients being real
smooth functions on $U$. Namely, every element of $\cA$ is of the
form $\sum_{i\ge 0} f_i\barh^i$ where $f_i$ are smooth functions on
$U$. Then $\cA$ is an $\R[[\barh]]$-module in the obvious way.

Given any two smooth functions $u$ and $v$ on $N^{1,n-1}$, we denote
by $u v$ the usual point-wise product of the two functions. We also
define their star-product (or more precisely, Moyal product) $u\ast
v$ on $U$ by
\begin{eqnarray}\label{multiplication}
(u\ast v)(x) = \lim_{x'\rightarrow x}\ \exp{\left(\barh \sum_{i j}
\theta_{i j}\partial_i\partial_j^\prime\right)}u(x) v(x'),
\end{eqnarray}
where $\partial_i=\frac{\partial}{\partial_i}$,  and $(\theta_{i
j})$ is a constant skew symmetric $n\times n$ matrix. In order for
the exponential to be dimensional less, the components $\theta_{i
j}$ of the matrix may need to have different dimensions. It is well
known that such a multiplication is associative. Since $\theta$ is
constant, the Leibniz rule remains valid in the present case:
\[ \partial_i(u\ast v)= \partial_i u\ast v + u\ast \partial_i v. \]

For positive integer $m=p+q$, we define a dot-product
\begin{eqnarray}\label{dotproduct-dimm}
 \bullet: \cA^m\otimes_{\R[[\barh]]} \cA^m\longrightarrow
\cA^m
\end{eqnarray}
over $U$ by
\begin{eqnarray*}
A\bullet B=-\sum _{i=1} ^p a_i\ast b_i +\sum _{j=p+1} ^{p+q} a_i
\ast b_i
\end{eqnarray*}
for all $A=(a_1, \dots, a_m)$ and $B=(b_1, \dots, b_m)$ in $\cA^m$.
The dot-product is a map of two-sided $\cA$-modules.

For a given $X\in \cA^m$, we let $E_i=\partial_i X$, and define
\begin{eqnarray*}
\g _{i j}=E_i \bullet E_j.
\end{eqnarray*}
Denote by $\g=(\g_{i j})$ the $n\times n$ matrix with entries $\g_{i
j}\in \cA$. If $\g \mod \barh$ is invertible over $U$, we
shall call $\g$ the {\em local quantum deformation of spacetime
metric} $g$ over $U$.

The discussion on the metric in \cite{CTZZ} carries over to the
present situation; in particular, the invertibility of $\g \mod
\barh$ implies that there exists a unique inverse $(\g^{i j})$ such
that
\begin{eqnarray*}
\g _{ij} \ast \g ^{jk} =\g ^{k j}  \ast \g _{j i} =\delta _i^k.
\end{eqnarray*}
Now as in \cite{CTZZ}, we define the left tangent bundle $TX$
(respectively right tangent bundle $\tilde TX$) of the local
noncommutative spacetime $(U, \g)$ as the left (respectively right)
$\cA$-submodule of $\cA^m$ generated by the elements $E_i$. The fact
that the metric $\g$ belongs to $GL_n(\cA)$ enables us to show that
the left and right tangent bundles are projective $\cA$-modules. Let
\begin{eqnarray*}\label{E-upper}
E^i = \g^{i j}\ast E_j, \quad \tilde E^i = E_j \ast \g^{j i},
\end{eqnarray*}
which belong to $TX$ and $\tilde TX$ respectively. Then the metric
gives rise to a $\cA$-bimodule map $\g:
TX\otimes_{\R[[\barh]]}\tilde TX \longrightarrow \cA$ with the
property
\[
\begin{aligned}
&\g(E_i, E_j) = \g_{i j}, \quad \g(E^i, \tilde E^j)= \g^{i j},\\
 &\g(E^i, E_j)=
\delta^i_j=\g(E_j, \tilde E^i).
\end{aligned}
\]

The connection $\nabla_i$ on the left tangent bundle will be defined
in the same way as \cite{CTZZ}, namely, by the composition of the
derivative $\partial_i$ with the projection from the free left
$\cA$-module $\cA^m$ onto the left tangent bundle. The connection
$\tilde\nabla_i$ on the right tangent bundle is defined similarly.
In order to describe the connections more explicitly, we note that
there exist $\Gamma_{i j}^k$ and $\tilde\Gamma_{i j}^k$ in $\cA$
such that
\begin{eqnarray}\label{Gamma}
\nabla_i E_j= \Gamma_{i j}^k \ast E_k, && \tilde \nabla_i E_j= E_k
\ast \tilde \Gamma_{i j}^k.
\end{eqnarray}
Because the metric is invertible, the elements $\Gamma_{i j}^k$ and
$\tilde \Gamma_{i j}^k$ are uniquely defined by equation
\eqref{Gamma}. We have
\begin{eqnarray}\label{Gamma1}
\Gamma_{i j}^k =  \partial_i E_j\bullet \tilde E^k &\quad&
\tilde\Gamma_{i j}^k = E^k \bullet  \partial_i E_j.
\end{eqnarray}
It is evident that $\Gamma_{i j}^k$ and $\tilde \Gamma_{i j}^k$ are
symmetric in the indices $i$ and $j$.  The following closely related
objects will also be useful later:
\[
\Gamma_{i j k} =  \partial_i E_j\bullet E_k , \quad
\tilde\Gamma_{i j k} = E_k \bullet  \partial_i E_j.
\]
In contrast to the commutative case, $\Gamma_{i j}^k$ and
$\tilde\Gamma_{i j}^k$ do not coincide in general. We have
\[
\Gamma_{i j}^k =  {}_c\Gamma_{i j l}\ast \g^{l k} + \Upsilon_{i j
 l}\ast \g^{l k}, \quad \tilde\Gamma_{i j}^k = \g^{k l}\ast
{}_c\Gamma_{i j l} - \g^{k l}\ast \Upsilon_{i j l},
\]
where
\begin{eqnarray*}
\begin{aligned}
_c\Gamma_{i j l} &= \frac{1}{2} \left(\partial_i \g_{j l} +
\partial_j \g_{l i}
-\partial_l \g_{j i} \right), \\
\Upsilon_{i j l} &= \frac{1}{2} \left(\partial_iE_j\bullet E_l -
E_l\bullet \partial_i E_j\right).
\end{aligned}
\end{eqnarray*}
Therefore the left and right connections involve two parts. The part
$ _c\Gamma_{i j l}$ depends on the metric only, while the {\em
noncommutative torsion} $\Upsilon_{i j l}$ embodies extra
information. In the present case, the noncommutative torsion depends
explicitly on the embedding. In the classical limit with $\barh=0$,
$\Upsilon_{i j}^k$ vanishes and both $\Gamma_{i j}^k$ and
$\tilde\Gamma_{i j}^k$ reduce to the standard Levi-Civita
connection.

The connections are metric compatible in the following sense
\cite[Proposition 2.7]{CTZZ}
\begin{eqnarray}\label{compatible}
\partial_i \g(Z, \tilde Z)=\g(\nabla_i Z, \tilde Z) + \g(Z, \tilde\nabla_i\tilde
Z), &\quad& \forall Z\in TX, \ \tilde Z\in \tilde TX.
\end{eqnarray}
This is equivalent to the fact that
\begin{eqnarray}\label{compatible1}
\partial_i \g_{j k} -\Gamma_{i j k} - \tilde\Gamma_{i k j} =0.
\end{eqnarray}

In contrast to the commutative case, equation \eqref{compatible1} by
itself is not sufficient to uniquely determine the connections
$\Gamma_{i j k}$ and $\tilde\Gamma_{i j k}$; the noncommutative
torsion needs to be specified independently.

Let $[\nabla_i, \nabla_j]:=\nabla_i \nabla_j- \nabla_j \nabla_i$ and
${[\tilde\nabla_i,  \tilde\nabla_j]}:= \tilde\nabla_i
\tilde\nabla_j- \tilde\nabla_j \tilde\nabla_i$. Straightforward
calculations show that for all $f\in \cA$,
\begin{eqnarray*}
\begin{aligned}
&{[}\nabla_i, \nabla_j{]}
(f\ast Z) =f\ast[\nabla_i, \nabla_j]Z,   &Z\in TX,  \\
&{[}\tilde\nabla_i,  \tilde\nabla_j{]}(W\ast f) = [\tilde\nabla_i,
\tilde\nabla_j]W\ast f, & W\in \tilde TX.
\end{aligned}
\end{eqnarray*}
Clearly the right-hand side of the first equation belongs to $TX$,
while that of the second equation belongs to $\tilde TX$. Thus The
maps $[\nabla_i, \nabla_j]$ and $[\tilde\nabla_i, \tilde\nabla_j]$
are left and right $\cA$-module homomorphisms respectively. So we
can always write
\begin{eqnarray}\label{curvature}
{[\nabla_i, \nabla_j]} E_k = R_{k i j}^l\ast E_l, & \quad &
{[}\tilde\nabla_i, \tilde\nabla_j{]}E_k = E_l\ast \tilde R_{k i
j}^l
\end{eqnarray}
for some $R_{k i j}^l, \tilde R_{k i j}^l\in\cA$.
We refer to $R_{k i j}^l$ and $\tilde R_{k i j}^l$ respectively as
the {\em Riemann curvatures} of the left and right tangent bundles
of the noncommutative spacetime $(U, \g)$. We have
\[
\begin{aligned} R_{k i j}^l &=-\partial_j\Gamma_{i k}^l -
\Gamma_{i k}^p \ast \Gamma_{j p}^l + \partial_i\Gamma_{j k}^l
+\Gamma_{j k}^p\ast \Gamma_{i p}^l ,\\
\tilde R_{k i j}^l &=-\partial_j\tilde\Gamma_{i k}^l -
\tilde\Gamma_{j p}^l\ast\tilde\Gamma_{i k}^p
+\partial_i\tilde\Gamma_{j k}^l + \tilde\Gamma_{i p}^l\ast
\tilde\Gamma_{j k}^p.
\end{aligned}
\]
Let us define
\begin{eqnarray*}
R_{l k i j} = R_{k i j}^p\ast \g_{p l}, &\quad& \tilde R_{l k i j} =
- \g_{k p}\ast \tilde R_{l i j}^p.
\end{eqnarray*}
Then these Riemannian curvatures of the left and right tangent
bundles coincide \cite[Lemma 2.12]{CTZZ}
\begin{eqnarray*}
R_{k l i j}=\tilde R_{ k l i j}.
\end{eqnarray*}
Therefore we only need to study the Riemannian curvature on one of
the tangent bundles.

Another important property of the Riemannian curvature is that it
satisfies the noncommutative analogues of the first and second
Bianchi identities \cite[Theorem 4.3]{CTZZ}. Note also that $ R_{ k
l i j}=-R_{k l j i }, $ but there is no simple rule to relate $R_{l
k i j }$ to $R_{k l i j}$ in contrast to the commutative case.

Let
\begin{eqnarray}
R_{i j} = R^p_{i p j}, &\quad& R= \g^{j i}\ast R_{i j},
\end{eqnarray}
and call them the {\em Ricci curvature} and {\em scalar curvature}
respectively. Let
\begin{eqnarray}
R^i_j = \g^{i k}\ast R_{k j},
\end{eqnarray}
then the scalar curvature is $R=R^i_i$. Let us also introduce the
following object
\begin{eqnarray}\label{Theta}
\Theta^l_p := \g^{i k}\ast R^l_{k p i},
\end{eqnarray}
which we will refer to as the $\Theta$-curvature.  In the
commutative case, $\Theta^l_p$ coincides with $R^l_p$, but it is no
longer true in the present setting. However, note that
\begin{eqnarray}\label{Scalar2}
\Theta^i_i=\g ^{ik} \ast R ^l _{kli} = \g ^{ik} \ast R _{ki} =R.
\end{eqnarray}

Analysis of the noncommutative second Bianchi identity in
\cite[\S4.B]{CTZZ} showed that $R^i_j + \Theta^i_j- \delta^i_j R$
was an analogue of the usual Einstein tensor (more precisely, $2$
times the Einstein tensor) in the case of vanishing cosmological
constant. Based on the analysis, a noncommutative Einstein equation
in the vacuum without cosmological constant was proposed in
\cite[\S4.C]{CTZZ}.

In view of \cite[\S4.B]{CTZZ}, the following appears to be a
reasonable proposal for a noncommutative Einstein equation over $U$,
\begin{eqnarray} \label{Einstein}
R^i_j + \Theta^i_j- \delta^i_j R +2\delta ^i _j \Lambda =2T^i _j,
\end{eqnarray}
where $T^i _j$ is some generalized ``energy-momentum tensor",  and
$\Lambda$ is the cosmological constant. This reduces to the vacuum
equation suggested in \cite{CTZZ} when $T^i_j=0$ and the
cosmological constant vanishes. We hope to provide a mathematical
justification for this proposal in future work, where the defining
properties of $T^i _j$ will also be specified.

In the commutative limit, we  recover the usual Einstein equation
from equation \eqref{Einstein}. However, it is yet to be seen
whether the equation in the noncommutative setting correctly
describes physics.

We shall test aspects of the validity of the noncommutative Einstein
equation by examining whether it is possible to find solutions of
the equation, which may be considered as noncommutative analogues of
physically important spacetimes in the commutative setting, for
example, the Schwarzschild spacetime and Schwarzschild-de Sitter
spacetime. As we shall see presently, this is indeed possible.

\section{Quantum deformation of the Schwarzschild spacetime}\label{schwarzschild}

In this section, we investigate noncommutative analogues of the
Schwarzschild spacetime using the general theory discussed in the
previous section.  Recall that the Schwarzschild spacetime has the
following metric
\begin{eqnarray}\label{Sch}
g=-\left(1-\frac{2m}{r}\right)dt^2
+\left(1-\frac{2m}{r}\right)^{-1}dr^2 +r^2\left(d\theta ^2 +\sin^2
\theta d\phi ^2 \right)
\end{eqnarray}
where $m=\frac{2 G M}{c^2}$ is constant, with $M$ interpreted as the
total mass of the spacetime. In the formula for $m$, $G$ is the
Newton constant, and $c$ is the speed of light. The Schwarzschild
spacetime can be embedded into a flat space of 6-dimensions in the
following two ways \cite{K2, F}:

\medskip
\noindent(i). {Kasner's embedding:}
\begin{eqnarray*}
\begin{aligned}
X^1&=\left(1-\frac{2m}{r}\right)^\frac{1}{2} \sin t,\\
X^2&=\left(1-\frac{2m}{r}\right)^\frac{1}{2} \cos t,\\
X^3&=f(r),\quad (f')^2 +1=\left(1-\frac{2m}{r}\right)^{-1}
\left(1+\frac{m^2}{r^4}\right),\\
X^4&=r \sin\theta \cos\phi,\\
X^5&=r \sin\theta \sin\phi,\\
X^6&=r \cos\theta,
\end{aligned}
\end{eqnarray*}
with the Schwarzschild metric given by
\begin{eqnarray*}
g=-\big(dX^1\big)^2-\big(dX^2\big)^2+\big(dX^3\big)^2+\big(dX^4\big)^2
+\big(dX^5\big)^2+\big(dX^6\big)^2
\end{eqnarray*}

\medskip
\noindent(ii). {Fronsdal's embedding:}
\begin{eqnarray*}
\begin{aligned}
Y^1&=\left(1-\frac{2m}{r}\right)^\frac{1}{2} \sinh t,\\
Y^2&=\left(1-\frac{2m}{r}\right)^\frac{1}{2} \cosh t,\\
Y^3&=f(r),\quad (f')^2 +1=\left(1-\frac{2m}{r}\right)^{-1}
\left(1-\frac{m^2}{r^4}\right),\\
Y^4&=r \sin\theta \cos\phi,\\
Y^5&=r \sin\theta \sin\phi,\\
Y^6&=r \cos\theta,
\end{aligned}
\end{eqnarray*}
with the Schwarzschild metric given by
\begin{eqnarray*}
g=-\big(dY^1\big)^2+\big(dY^2\big)^2+\big(dY^3\big)^2+\big(dY^4\big)^2
+\big(dY^5\big)^2+\big(dY^6\big)^2.
\end{eqnarray*}

Let us now construct a noncommutative analogue of the Schwarzschild
spacetime. Denote $x^0=t$, $x^1=r$, $x^2=\theta$ and $x^3=\phi$. We
deform the algebra of functions in these variables by imposing on it
the Moyal product defined by \eqref{multiplication} with the
following anti-symmetric matrix
\begin{eqnarray}
\left(\theta _{\mu \nu}\right)_{\mu,
\nu=0}^3=\left(\begin{array}{cccc}
   0&  0&  0&  0\\
   0&  0&  0&  0\\
   0&  0&  0&  1\\
   0&  0&  -1&  0
\end{array}\right).\label{asy-matrix}
\end{eqnarray}
Denote the resultant noncommutative algebra by $\cA$. Note that in
the present case that the nonzero components of the matrix
$\left(\theta _{\mu \nu}\right)$ are dimensionless.

Now we regard the functions $X^i$, $Y^i$  $(1\le i\le 6)$ appearing
in both Kasner's and Fronsdal's embeddings as elements of $\cA$. For
$\mu=0, 1, 2, 3$, and $i=1, 2, \dots, 6$, let
\begin{eqnarray}\label{veibein}
\begin{aligned}
E^i_\mu&=& \frac{\partial X^i}{\partial x^\mu},
&\quad \text{for Kasner's embedding}, \\
E^i_\mu&=& \frac{\partial Y^i}{\partial x^\mu},
&\quad \text{for
Fronsdal's embedding}.
\end{aligned}
\end{eqnarray}
Following the general theory of the last section, we define the
metric and noncommutative torsion for the noncommutative
Schwarzschild spacetime by,

\medskip
\noindent (1). in the case of Kasner's embedding
\begin{eqnarray}\label{Kasner-case}
\begin{aligned}
\g_{\mu \nu}=& -E^1_\mu\ast E^1_\nu - E^2_\mu\ast E^2_\nu +
\sum_{j=3}^6 E^j_\mu\ast E^j_\nu,\\
\Upsilon_{\mu\nu\rho} =& \frac{1}{2}\left(-\partial_\mu E^1_\nu\ast
E^1_\rho -  \partial_\mu E^2_\nu\ast E^2_\rho +\sum_{j=3}^6
\partial_\mu E^j_\nu\ast
E^j_\rho \right)\\
&+\frac{1}{2}\left(-E^1_\rho\ast  \partial_\mu E^1_\nu -E^2_\rho\ast
\partial_\mu E^2_\nu+\sum_{j=3}^6
E^j_\rho  \ast \partial_\mu E^j_\nu\right);
\end{aligned}
\end{eqnarray}
\noindent (2). in the case of Fronsdal's embedding
\begin{eqnarray}\label{Fronsdal-case}
\begin{aligned}
\g_{\mu \nu}=& -E^1_\mu\ast E^1_\nu + \sum_{j=2}^6 E^j_\mu\ast
E^j_\nu, \\
\Upsilon_{\mu\nu\rho} =& \frac{1}{2}\left(-\partial_\mu E^1_\nu\ast
E^1_\rho +\sum_{j=2}^6
\partial_\mu E^j_\nu\ast
E^j_\rho \right)\\
&+\frac{1}{2}\left(-E^1_\rho\ast  \partial_\mu E^1_\nu +\sum_{j=2}^6
E^j_\rho  \ast \partial_\mu E^j_\nu\right).
\end{aligned}
\end{eqnarray}

Some lengthy but straightforward calculations show that the metrics
and the noncommutative torsions are respectively equal in the two
cases. Since the noncommutative torsion will not be used in later
discussions, we shall not spell it out explicitly. However, we
record the metric $\g=(\g_{\mu \nu})$ of the quantum deformation of
the Schwarzschild spacetime below:
\begin{eqnarray}\label{deformed-Sch}
\begin{aligned}
\g_{0 0} =&-\left(1-\frac{2m}{r}\right),\\
\g_{0 1} =&\g_{1 0}=\g_{0 2} =\g_{2 0}=\g_{0 3} =\g_{3 0}=0, \\
\g_{1 1}=&\left(1-\frac{2m}{r}\right)^{-1}
\left[1+\left(1-\frac{2m}{r}\right)
          \left(\sin^2\theta -\cos^2 \theta\right)\sinh^2\barh\right],\\
\g_{1 2} =    &\g_{2 1} = 2r\sin\theta\cos\theta\sinh^2\barh,\\
\g_{1 3} =    &-\g_{3 1} = -2r\sin\theta\cos\theta\sinh\barh\cosh\barh,\\
\g_{2 2} =    &r^2
    \left[1-\left(\sin^2\theta-\cos^2\theta\right)\sinh^2\barh\right],\\
\g_{2 3} = &-\g_{3 2}
=r^2\left(\sin^2\theta-\cos^2\theta\right)\sinh\barh\cosh\barh,\\
\g_{3 3} =&r^2
    \left[\sin^2\theta+\left(\sin^2\theta-\cos^2\theta\right)\sinh^2\barh\right].
\end{aligned}
\end{eqnarray}

It is interesting to observe that the quantum deformation of the
Schwarzschild metric (\ref{deformed-Sch}) still has a black hole
with the event horizon at $r=2m$. The Hawking temperature and
entropy of the black hole are respectively given by
\begin{eqnarray*}
\begin{aligned}
T=\frac{1}{2}\frac{d\g_{00}}{dr}\Big|_{r=2m}=\frac{1}{4m},\qquad
S_{bh}=4\pi m^2.
\end{aligned}
\end{eqnarray*}
They coincide with the temperature and entropy of the classical
Schwarzschild black hole of mass $M$. However, the area of the event
horizon of the noncommutative black hole receives corrections from
the quantum deformation of the spacetime. Let
$\bar{g}=\begin{pmatrix}\g_{2 2} & \g_{2 3}\\ \g_{3 2} & \g_{3 3}
\end{pmatrix}.$ We have
\begin{eqnarray*}
\begin{aligned}
A=&\iint _{\{r=2m\}} \sqrt{\det\bar{g}}\, d\theta d\phi\\
=&\iint _{\{r=2m\}} r^2 \sin\theta \sqrt{1+\left(\sin^2\theta
-\cos^2\theta\right)\sinh^2 \barh}d\theta d\phi\\
=&16\pi m^2 \left(1-\frac{\barh^2}{6} +O(\barh^4)\right).
\end{aligned}
\end{eqnarray*}
This leads to the following relationship between the horizon area
and entropy of the noncommutative black hole:
\begin{eqnarray}\label{entropy}
S_{bh} =\frac{A}{4}\left(1+\frac{\barh^2}{6} +O(\barh^4)\right).
\end{eqnarray}

Let us now consider the Ricci and $\Theta$-curvature of the deformed
Schwarzschild metric. We have
\begin{eqnarray}\label{RTheta}
\begin{aligned}
R_0^1=&R_0^2=R_0^3=R_1^0=R_2^0=R_3^0=0,\\
\Theta_0^1=&\Theta_0^2=\Theta_0^3=\Theta_1^0=\Theta_2^0=\Theta_3^0=0,\\
R_0 ^0 =&\Theta _0 ^0=-\frac{m\left[ 2m + 3r + 3\left( m + r
\right)\cos2\theta\right]}{r^4}\barh^2 +O(\barh^4),\\
R_1 ^1 =&\Theta _1 ^1=\frac{m\left[-14m + 3r + \left( -11m + r
\right)\cos2\theta\right]}{r^4} \barh^2+ O(\barh^4),\\
R_1 ^2 =&\Theta _1 ^2=\frac{2m\cos^2\theta \cot\theta}{r^4} \barh^2+
O(\barh^4),\\
R_1 ^3 =&-\Theta _1 ^3=\frac{2m\cot\theta}{r^4}\barh +O(\barh^3),\\
R_2 ^1 =&\Theta _2 ^1=\frac{5m\left( -2m + r \right)\sin
2\theta}{r^3} \barh^2+ O(\barh^4),\\
R_2 ^2 =&\Theta _2 ^2=\frac{m\left[4\left(m + r \right)  + \left( 6m
+ 5r \right)\cos2\theta \right]}{r^4}\barh^2 + O(\barh^4),\\
R_2 ^3 =&-\Theta _2 ^3=\frac{4m}{r^3}\barh +O(\barh^3),\\
R_3 ^1 =&-\Theta _3 ^1=\frac{m\left( 2m - r
\right)\sin2\theta}{r^3}\barh+O(\barh^3).\\
R_3 ^2 =&-\Theta _3 ^2=\frac{4m \cos^2\theta}{r^3}\barh+O(\barh^3).\\
R_3 ^3 =&\Theta _3 ^3=\frac{m\left[-8m +8r+\left( -6m + 9r \right)
\cos2\theta \right]}{r^4}\barh^2 + O(\barh^4).
\end{aligned}
\end{eqnarray}

Note that $R_i^i=\Theta_i^i$ for all $i$, and $R_i^j=-\Theta_i^j$ if
$i\ne j$. Let us write
\begin{eqnarray}\label{expansion}
\begin{aligned}
R_j^i= {R_j^i}_{(0)} + \barh {R_j^i}_{(1)} + \barh^2
{R_j^i}_{(2)}+\dots, \\
\Theta_j^i= {\Theta_j^i}_{(0)} + \barh {\Theta_j^i}_{(1)} + \barh^2
{\Theta_j^i}_{(2)}+\dots. \end{aligned}
\end{eqnarray}
Then the formulae for $R_j^i$ and $\Theta_j^i$ show that
\[
{R_j^i}_{(0)}={\Theta_j^i}_{(0)}, \quad
{R_j^i}_{(1)}=-{\Theta_j^i}_{(1)}, \quad {R_j^i}_{(2)}
={\Theta_j^i}_{(2)}.
\]

Naively generalizing the Einstein tensor $R^i _j-
\frac{1}{2}\delta^i_j R$ to the noncommutative setting, one ends up
with a quantity that does not vanish at order $\barh$, as can be
easily shown using the above results. However,
\[ R^i _j +\Theta ^i _j
-\delta ^i _j R =0 + O(\barh^2).
\]
This indicates that the proposed noncommutative Einstein equation
\eqref{Einstein} captures some essence of the underlying symmetries
in the noncommutative world.

Now the deformed Schwarzschild metric (\ref{deformed-Sch}) satisfies
the vacuum noncommutative Einstein equation (\ref{Einstein}) with
$T^i _j=0$ and $\Lambda=0$ to first order in the deformation
parameter. However, if we take into account higher order corrections
in $\barh$, the deformed Schwarzschild metric no longer satisfies
the noncommutative Einstein equation in the vacuum. Instead, $R^i _j
+\Theta ^i _j -\delta ^i _j R = T^i_j$ with $T^i_j$ being of order
$O(\barh^2)$ and given by
\begin{eqnarray}\label{T}
\begin{aligned}
T_0^1=&T_0^2=T_0^3=T_1^0=T_2^0=T_3^0=T_3^1=T_3^2=T_1^3=T_2^3=0,\\
T_0^0=&\frac{m\left[ 8m - 9r + \left( 4m - 9r
\right)\cos2\theta\right]}{r^4}\barh^2 + O(\barh^4),\\
T_1^1=&\frac{-m\left[ 4m + 3r + \left( 4m + 5r
\right)\cos2\theta\right]}{r^4}\barh^2 + O(\barh^4),\\
T_1^2=&\frac{2m \cos^2 \theta \cot\theta }{r^4}\barh^2 +O(\barh^4),\\
T_2^1=&\frac{5m\left( -2m + r \right) \sin2\theta}{r^3}\barh^2 + O(\barh^4),\\
T_2^2=&\frac{m\left[ 14m - 2r + \left( 13m - r
\right)\cos2\theta\right]}{r^4}\barh^2 + O(\barh^4),\\
T_3^3=&\frac{m\left[2\left(m + r \right)  + \left( m + 3r
\right)\cos2\theta\right]}{r^4}\barh^2 +O(\barh^4).
\end{aligned}
\end{eqnarray}

A possible physical interpretation of the results is the following.
We regard the $\barh$ and higher order terms in the metric $\g_{i
j}$ and associated curvature $R_{i j k l}$ as arising from quantum
effects of gravity. Then the $T_i^j$ obtained in \eqref{T} should be
interpreted as quantum corrections to the classical Einstein tensor.

\section{Quantum deformation of the Schwarzschild-de
Sitter spacetime}\label{schwarzschild-dS}

In this section, we investigate a noncommutative analogue of the
Schwarzschild-de Sitter spacetime. Since the analysis is parallel to
that on the quantum Schwarzschild spacetime, we shall only present
the pertinent results.

Recall that the Schwarzschild-de Sitter spacetime has the following
metric
\begin{eqnarray}\label{Sch-dS}
g=-\left(1-\frac{r^2}{l^2}-\frac{2m}{r}\right)dt^2
+\left(1-\frac{r^2}{l^2}-\frac{2m}{r}\right)^{-1}dr^2
+r^2\left(d\theta ^2 +\sin^2 \theta d\phi ^2 \right)
\end{eqnarray}
where $\frac{3}{l^2}=\Lambda>0$ is the cosmological constant, and
$m$ is related to the total mass of the spacetime through the same
formula as in the Schwarzschild case. This spacetime can be embedded
into a flat space of 6-dimensions in two different ways.

\medskip
\noindent(i). {Generalized Kasner embedding:}
\begin{eqnarray*}
\begin{aligned}
X^1&=\left(1-\frac{r^2}{l^2}-\frac{2m}{r}\right)^\frac{1}{2} \sin t,\\
X^2&=\left(1-\frac{r^2}{l^2}-\frac{2m}{r}\right)^\frac{1}{2} \cos t,\\
X^3&=f(r),\quad
(f')^2+1=\left(1-\frac{r^2}{l^2}-\frac{2m}{r}\right)^{-1}
\left[1+\left(\frac{m}{r^2}-\frac{r}{l^2}\right)^2\right],\\
X^4&=r \sin\theta \cos\phi,\\
X^5&=r \sin\theta \sin\phi,\\
X^6&=r \cos\theta,
\end{aligned}
\end{eqnarray*}
with the Schwarzschild-de Sitter metric given by
\begin{eqnarray*}
g=-\big(dX^1\big)^2-\big(dX^2\big)^2+\big(dX^3\big)^2+\big(dX^4\big)^2
+\big(dX^5\big)^2+\big(dX^6\big)^2
\end{eqnarray*}

\medskip
\noindent(ii). {Generalized Fronsdal embedding:}
\begin{eqnarray*}
\begin{aligned}
Y^1&=\left(1-\frac{r^2}{l^2}-\frac{2m}{r}\right)^\frac{1}{2} \sinh t,\\
Y^2&=\left(1-\frac{r^2}{l^2}-\frac{2m}{r}\right)^\frac{1}{2} \cosh t,\\
Y^3&=f(r),\quad (f')^2
+1=\left(1-\frac{r^2}{l^2}-\frac{2m}{r}\right)^{-1}
\left[1-\left(\frac{m}{r^2}-\frac{r}{l^2}\right)^2\right],\\
Y^4&=r \sin\theta \cos\phi,\\
Y^5&=r \sin\theta \sin\phi,\\
Y^6&=r \cos\theta,
\end{aligned}
\end{eqnarray*}
with the Schwarzschild-de Sitter metric given by
\begin{eqnarray*}
g=-\big(dY^1\big)^2+\big(dY^2\big)^2+\big(dY^3\big)^2+\big(dY^4\big)^2
+\big(dY^5\big)^2+\big(dY^6\big)^2.
\end{eqnarray*}

Let us now construct a noncommutative analogue of the
Schwarzschild-de Sitter spacetime. Denote $x^0=t$, $x^1=r$,
$x^2=\theta$ and $x^3=\phi$. We deform the algebra of functions in
these variables by imposing on it the Moyal product defined by
\eqref{multiplication} with the anti-symmetric matrix
(\ref{asy-matrix}).  Denote the resultant noncommutative algebra by
$\cA$.

Now we regard the functions $X^i$, $Y^i$  $(1\le i\le 6)$ appearing
in both the generalized Kasner embedding  and the generalized
Fronsdal embedding as elements of $\cA$. Let $E_\mu^i$ ($\mu=0, 1,
2, 3$, and $i=1, 2, \dots, 6$) be defined by \eqref{veibein} but for
the generalized Kasner and  Fronsdal embeddings respectively. We
also define the metric and noncommutative torsion for the
noncommutative Schwarzschild-de Sitter spacetime by equations
\eqref{Kasner-case} and \eqref{Fronsdal-case} for the generalized
Kasner and Fronsdal embeddings respectively. As in the case of the
noncommutative Schwarzschild spacetime, we can show that the metrics
and the noncommutative torsions are respectively equal for the two
embeddings. We record the metric $\g=(\g_{\mu \nu})$ of the quantum
deformation of the Schwarzschild-de Sitter spacetime below:
\begin{eqnarray}\label{deformed-Sch-dS}
\begin{aligned}
\g_{0 0} =&-\left(1-\frac{r^2}{l^2}-\frac{2m}{r}\right),\\
\g_{0 1} =&\g_{1 0}=\g_{0 2} =\g_{2 0}=\g_{0 3} =\g_{3 0}=0, \\
\g_{1 1}=&\left(1-\frac{r^2}{l^2}-\frac{2m}{r}\right)^{-1}
\left[1+\left(1-\frac{r^2}{l^2}-\frac{2m}{r}\right)
          \left(\sin^2\theta -\cos^2 \theta\right)\sinh^2\barh\right],\\
\g_{1 2} =    &\g_{2 1} = 2r\sin\theta\cos\theta\sinh^2\barh,\\
\g_{1 3} =    &-\g_{3 1} = -2r\sin\theta\cos\theta\sinh\barh\cosh\barh,\\
\g_{2 2} =    &r^2
    \left[1-\left(\sin^2\theta-\cos^2\theta\right)\sinh^2\barh\right],\\
\g_{2 3} = &-\g_{3 2}
=r^2\left(\sin^2\theta-\cos^2\theta\right)\sinh\barh\cosh\barh,\\
{\g}_{3 3} =&r^2
    \left[\sin^2\theta+\left(\sin^2\theta-\cos^2\theta\right)\sinh^2\barh\right].
\end{aligned}
\end{eqnarray}

Let us now consider the Ricci curvature and the $\Theta$-curvature
of the deformed Schwarzschild metric. We have
\begin{eqnarray}\label{RTheta-dS}
\begin{aligned}
R_0^1=&R_0^2=R_0^3=R_1^0=R_2^0=R_3^0=0,\\
\Theta_0^1=&\Theta_0^2=\Theta_0^3=\Theta_1^0=\Theta_2^0=\Theta_3^0=0,\\
R_0^0=&\Theta_0^0=\frac{3}{l^2} + \Big[ l^2 \left( 10m - 3r
\right)r^3 + 10r^6
- l^4 m \left( 2m + 3r \right)\\
&-3\left\{ -2 l^2 m r^3 - 2 r^6 + l^4 m
\left( m + r \right)  \right\} \cos 2\theta \Big] \frac{\barh^2}{l^4 r^4}
+O(\barh^4),\\
R_1^1=&\Theta_1^1=\frac{3}{l^2} + \Big[ l^2 \left( 16 m - 9r \right)
r^3 + 16
r^6 + l^4 m \left( -14 m + 3 r \right) \\
& + \left\{ 2 l^2 \left( 5 m - 2 r \right) r^3 + 10 r^6 + l^4 m
\left( -11 m + r \right) \right\}
\cos 2\theta \Big] \frac{\barh^2}{l^4 r^4} +O(\barh^4),\\
R_1^2=&\Theta_1^2=\frac{2 \left( l^2 m - 4 r^3 \right)
\cos^2 \theta \cot \theta}{l^2 r^4} \barh^2 +O(\barh^4),\\
R_1^3=&-\Theta_1^3=\frac{2\, \left(l^2 m - 4r^3 \right)
\cot\theta}{l^2 r^4}\barh +O(\barh^3),\\
R_2^1=&\Theta_2^1=-\frac{\left[ l^2 \left( 2m - r \right)  + r^3
\right]\left(5 l^2 m + 4 r^3 \right) \sin 2\theta}{l^4 r^3}\barh^2 +O(\barh^4),\\
R_2^2=&\Theta_2^2=\frac{3}{l^2} + \Big[ l^2 \left( 22 m - r \right)
r^3 + 10
r^6 + 4 l^4 m \left(m + r \right) \\
& + \left\{ 6r^6 + l^2 r^3 \left(15 m + 4 r \right)  + l^4 m \left(
6m + 5r \right)  \right\}\cos 2\theta \Big] \frac{\barh^2}{l^4 r^4} +O(\barh^4),\\
R_2^3=&-\Theta_2^3=\left( \frac{8}{l^2} + \frac{4m}{r^3} \right) \barh +O(\barh^3),\\
R_3^1=&-\Theta_3^1=\frac{\left( l^2 m - 4r^3 \right) \left[l^2 \left( 2m - r \right)
                   + r^3 \right] \sin 2\theta}{l^4 r^3} \barh+O(\barh^3),\\
R_3^2=&-\Theta_3^2=4 \left( \frac{2}{l^2} + \frac{m}{r^3} \right) \,
\cos^2\theta \barh +O(\barh^3),\\
R_3^3=&\Theta_3^3=\frac{3}{l^2} + \Big[ -8 l^4 m \left( m - r
\right) +l^2 \left( 28 m - 5r \right)r^3 + 16 r^6 \\
              &+ 3\left\{ 7 l^2 m r^3 + 4 r^6 + l^4 m
              \left( -2 m + 3r \right)  \right\} \cos 2\theta \Big]
              \frac{\barh^2}{l^4 r^4}
              +O(\barh^4).
\end{aligned}
\end{eqnarray}
Note that if we expand $R^j_i$ and $\Theta^j_i$ into power series in
$\barh$ in the form \eqref{expansion}, we again have
\[
{R_j^i}_{(0)}={\Theta_j^i}_{(0)}, \quad
{R_j^i}_{(1)}=-{\Theta_j^i}_{(1)}, \quad {R_j^i}_{(2)}
={\Theta_j^i}_{(2)}.
\]

By using the above results one can easily show that the deformed
Schwarzschild-de Sitter metric (\ref{deformed-Sch-dS}) satisfies the
vacuum noncommutative Einstein equation (\ref{Einstein}) (with $T^i
_j=0$) to first order in the deformation parameter:

\[ R^i _j +\Theta ^i _j
-\delta ^i _j R +\delta^i _j \frac{6}{l^2}=0 + O(\barh^2).
\]

Further analysing the deformed Schwarzschild-de Sitter metric, we
note that $R^i _j +\Theta ^i _j -\delta ^i _j R +\delta^i _j
\frac{6}{l^2}=T^i_j$ with $T^i_j$ being of order $O(\barh^2)$ and
given by
\begin{eqnarray*}
\begin{aligned}
T_0^1=&T_0^2=T_0^3=T_1^0=T_2^0=T_3^0=T_3^1=T_3^2=T_1^3=T_2^3=0,\\
T_0^0=&- \Big[ 2 l^2 \left( 14m-3r \right)r^3 + 16 r^6 + l^4 m
\left( -8m + 9r \right) \\
& + \left\{ 20 l^2 m r^3 + 11 r^6 + l^4 m \left( -4m + 9r \right)
\right\}\cos2\theta \Big]\frac{\barh^2}{l^4 r^4} +O(\barh^4),\\
T_1^1=&- \Big[ 22 l^2 m r^3 + 10 r^6 + l^4 m \left(4 m + 3r
\right)\\&+ \left\{7r^6 + 4 l^2 r^3 \left( 4 m + r \right)  + l^4 m
\left( 4 m+ 5 r \right)  \right\} \cos 2\theta \Big]
\frac{\barh^2}{l^4 r^4}+O(\barh^4),\\
T_1^2=&\frac{2 \left( l^2 m - 4 r^3 \right) \cos^2\theta \cot\theta}
{l^2 r^4}\barh^2 +O(\barh^4),\\
T_2^1=&-\frac{\left[ l^2 \left( 2m - r \right)  + r^3 \right]\left(
5 l^2m + 4 r^3 \right) \sin2\theta}{l^4 r^3} \barh^2+O(\barh^4),\\
T_2^2=&-\Big[2 \left\{ 4 l^2 \left( 2m - r \right) r^3 + 8 r^6 + l^4
m \left( -7m + r \right)  \right\} \\
& + \left\{l^2 \left(11m - 4r \right)r^3 + 11 r^6 + l^4 m \left( -13
m + r \right)\right\}\cos 2\theta \Big] \frac{\barh^2}{l^4 r^4} +O(\barh^4),\\
T_3^3=& \Big[2 \left\{ -5 r^6 + l^4 m \left( m + r \right) + l^2 r^3
\left( -5m + 2r \right)  \right\} \\
& + \left\{ -5 l^2 m r^3 - 5 r^6 + l^4 m \left( m + 3r \right)
\right\} \cos 2\theta \Big] \frac{\barh^2}{l^4 r^4}+O(\barh^4).
\end{aligned}
\end{eqnarray*}
Similar to the case of the quantum Schwarzschild spacetime, one may
regard this as quantum corrections to the Einstein tensor.

\section{Conclusion}\label{conclusion}

Working within the framework of the noncommutative Riemannian
geometry of \cite{CTZZ}, we have obtained in this paper quantum
analogues of the Schwarzschild spacetime and Schwarzschild-de Sitter
spacetime, and studied their noncommutative geometries. The quantum
Schwarzschild spacetime has been constructed in two ways,
respectively mimicking the Kasner and Fronsdal embeddings of the
classical Schwarzschild spacetime in $6$-dimensions. The metrics and
curvatures of the resultant quantum spacetimes coincide, and are
shown to be solutions of a noncommutative analogue of the Einstein
equation in the vacuum exact to the first order in the deformation
parameter. The Hawking temperature and entropy of the quantum
Schwarzschild black hole have been computed and shown to coincide
with the usual quantities. However, the area of the horizon has
received quantum corrections, and this in turn leads to a
modification of the entropy-area law. We have also constructed the
quantum Schwarzschild-de Sitter spacetime using two embeddings, and
shown that to the first order in the deformation parameter, the
spacetime is a solution of the vacuum noncommutative Einstein
equation with a cosmological constant.

Quantum deformations of the plane-fronted waves and other spacetimes
will be studied in a forthcoming paper \cite{WZZ}.

Works on noncommutative relativity and black holes reported in the
literature are largely based on physical intuitions. A fundamental
theory is much desired for developing the subject to a higher level
of sophistication. We hope that future work will develop the theory
of \cite{CTZZ} into a coherent mathematical framework for
noncommutative gravity.

Two problems deserve particular attention. One is a possible first
principle derivation of the noncommutative Einstein equation
\eqref{Einstein} (e.g., based on an action principle), the other is
the understanding of the symmetries governing \eqref{Einstein}. The
latter problem is closely related to the study of a noncommutative
analogue of the ``diffeomorphism group". In \cite[\S V]{CTZZ}, we
introduced noncommutative general coordinate transformations and
obtained the transformation rules of the metric, Riemannian
curvature tensor and other quantities under such transformations.
While the material of \cite[\S V]{CTZZ} is well adapted to studying
symmetries of equation \eqref{Einstein}, the approach of \cite{R7}
based on deforming the Hopf algebraic structure of differential
operators will also be worth investigating. We will return to the
issues alluded to here in future publications.

\bigskip

\bigskip

\noindent{\bf Acknowledgement}: We are indebted to M. Chaichian,  A.
Tureanu and M. Oksanen for valuable suggestions on the manuscript.
X. Zhang wishes to thank the School of Mathematics and Statistics,
University of Sydney for the hospitality during his visits when part
of this work was carried out. Partial financial support from the
Australian Research Council, National Science Foundation of China
(grants 10421001, 10725105, 10731080), NKBRPC (2006CB805905) and the
Chinese Academy of Sciences is gratefully acknowledged.

\end{document}